\documentclass[5p,times]{elsarticle}
\pdfoutput=1
\usepackage[numbers]{natbib}
\usepackage{float}
\usepackage{subfig}
\usepackage{wrapfig}
\usepackage{graphicx}
\usepackage{url}
\usepackage{xcolor}
\usepackage{stfloats}
\usepackage{tcolorbox}
\usepackage{amsmath}
\usepackage{amsfonts}
\usepackage{multirow}
\usepackage{booktabs}
\usepackage{listings}
\usepackage{pifont}
\usepackage{makecell}
\usepackage{nicematrix}
\usepackage{bm}
\usepackage{threeparttable}

\definecolor{verylightgray}{rgb}{.97,.97,.97}

\newcommand{\tool}{SCCLLM}

\newcommand{\RQone}{\textbf{RQ1: How effective is {\tool} when compared with state-of-the-art  baselines via automatic evaluation?}}

\newcommand{\RQtwo}{\textbf{RQ2: How effective is our proposed demonstration example selection strategy in  {\tool}?}}

\newcommand{\RQthree}{\textbf{RQ3: Whether the number of demonstration examples affect the effectiveness of {\tool}?}}

\newcommand{\RQfour}{\textbf{RQ4: How effective is  {\tool} when compared with state-of-the-art baselines via human study?}}

\journal{Information and Software Technology}

\begin{document}

\begin{frontmatter}
	
	\title{Automatic Smart Contract Comment Generation via Large Language Models and  In-Context Learning}

	\author[NTU]{Junjie Zhao}
	\ead{zhaojunjie225@gmail.com}	
        \author[NTU]{Xiang Chen\corref{mycorrespondingauthor}}
        \cortext[mycorrespondingauthor]{Corresponding author}
	\ead{xchencs@ntu.edu.cn}		
        \author[NUAA]{Guang Yang}
	\ead{novelyg@outlook.com}	
         \author[NTU]{Yiheng Shen}
	\ead{yiheng.s@outlook.com}	
 
	\address[NTU]{School of Information Science and Technology, Nantong University, Nantong, China}
	\address[NUAA]{College of Computer Science and Technology, Nanjing University of Aeronautics and Astronautics, Nanjing, China}

\begin{abstract}
\textbf{Context:}
Designing effective automatic smart contract comment generation approaches can facilitate developers’ comprehension, boosting
smart contract development and improving vulnerability detection.
The previous approaches can be divided into two categories: fine-tuning paradigm-based approaches and information retrieval-based approaches.

\noindent\textbf{Objective:}
However, for the fine-tuning paradigm-based approaches, the performance may be limited by the quality of the gathered dataset for the downstream task and they may have knowledge-forgetting issues,  which can reduce the generality of the fine-tuned model. While for the information retrieval-based approaches, it is difficult for them to generate high-quality comments if similar code does not exist in the historical repository.
Therefore we want to utilize the domain knowledge related to smart contract code comment generation in large language models (LLMs) to alleviate the disadvantages of these two types of approaches.

\noindent\textbf{Method:}
In this study, we propose an approach {\tool} based on LLMs and in-context learning.
Specifically, in the demonstration selection phase, {\tool} retrieves the top-$k$ code snippets from the historical corpus by considering syntax, semantics, and lexical information.
In the in-context learning phase, {\tool} utilizes the retrieved code snippets as demonstrations for in-context learning, which can help to utilize the related knowledge for this task in the LLMs.
In the LLMs inference phase, the input is the target smart contract code snippet, and the output is the corresponding comment generated by the LLMs.

\noindent\textbf{Results:}
We select a large corpus from a smart contract community Etherscan.io as our experimental subject. 
Extensive experimental results show the effectiveness of {\tool} when compared with baselines in automatic evaluation and human evaluation.
We also show the rationality of our customized demonstration selection strategy in {\tool} by ablation studies.

\noindent\textbf{Conclusion:}
Our study shows using LLMs and in-context learning is a promising direction for automatic smart contract comment generation, which calls for more follow-up studies.

\end{abstract}


\end{frontmatter}


\section{Introduction}
\label{sec:intro}

Smart contracts~\cite{zou2019smart,zheng2020overview} are self-executing digital contracts running on blockchain technology. They automate, validate, and enforce agreement terms without intermediaries, offering transparency and security.
However, Yang et al.~\cite{yang2021multi} found that most of the smart contract code comments are unavailable, which can make it challenging for developers to understand the code's logic, purpose, and intended functionality.
Moreover, smart contracts are susceptible to vulnerabilities and exploits.
In a previous study, He et al.~\cite{he2020characterizing} found that 10\% of the vulnerabilities were caused by code clones. If smart contract code lacks comments to explain potential risks and mitigate strategies, it becomes difficult to identify and address security vulnerabilities, which can increase the chances of hacks or attacks.
To this end, it is necessary to automatically generate concise and ﬂuent natural language descriptions for smart contract codes.
Based on the above analysis, we can find designing effective automatic comment generation approaches can facilitate developers' comprehension, boosting smart contract development and detecting vulnerabilities.
However, when compared to source code summarization~\cite{hu2020deep,li2022setransformer,li2021secnn}, the specific challenges associated with smart contract comment generation can be summarized as follows. First, smart contracts are typically written in Solidity. Understanding their codes requires specialized knowledge of these languages and the Ethereum platform. Second, given the implications of smart contracts in financial transactions and agreements, generating relevant and concise comments is crucial. Any misinterpretation could potentially have significant financial or legal implications. Finally, smart contracts encapsulate detailed business logic, which can be complex and multifaceted. This business logic should be adequately captured in the comments for a comprehensive understanding.

Until now, smart contract comment generation has received continuous attention. For example,
Yang et al.~\cite{yang2021multi} proposed the approach MMTrans. This approach learns the smart contract code representation from two heterogeneous modalities: SBT sequences~\cite{hu2018deep} (i.e., global semantic information) and graphs (i.e., local semantic information) based on abstract syntax trees. Later MMTrans uses two encoders to extract the semantic information from these two modalities respectively and then uses a joint decoder to generate code comments. 
Later we~\cite{yang2022ccgir} proposed an information retrieval-based approach CCGIR due to the widespread presence of code cloning in smart contract development. This approach employs CodeBert~\cite{feng2020codebert} to extract semantic vectors from the target code snippet and retrieve the top-$k$ code snippets based on their semantic similarity scores. Subsequently, it further considers the syntactic and lexical similarity of the code by combining these scores, which leads to the retrieval of the most similar code snippet. The comment from this retrieved code snippet is then reused for the target code snippet.

However, for the fine-tuning paradigm-based approaches~\cite{yang2021multi}, the performance may be limited by the quality of the gathered dataset for the downstream task.
Moreover, they may have the issue of knowledge forgetting~\cite{de2021continual}. Specifically, if a pre-trained model is fine-tuned on a specific downstream task, the model might start to forget the general knowledge it acquired during the pretraining phase. This issue can limit the model's generalization and versatility.
While for the information retrieval-based approaches~\cite{yang2022ccgir}, it is difficult for them to generate high-quality comments if similar smart contract codes do not exist in the historical repository.
To overcome the limitations of these two kinds of approaches for the smart contract comment generation, we want to leverage the emerging capabilities of large language models (LLMs), which have been pre-trained on vast amounts of data and possess a wealth of hidden domain knowledge, for automatic smart contract comment generation.
However, merely utilizing LLMs without effectively leveraging their related domain knowledge may not yield optimal results. 
As highlighted in Section~\ref{sec:resultRQ3}, our experiments reveal that directly employing LLMs in the zero-shot learning setting\footnote{For large language models, zero-shot learning refers to the capability of the model to perform tasks without explicit examples.} fails to outperform baselines based on fine-tuning.
To address this limitation, recent research has shown that the in-context learning paradigm offers a promising solution for harnessing the domain knowledge encapsulated within LLMs~\cite{nashid2023retrieval}. Specifically, given limited examples as the prompt, this paradigm can imitate the human ability to leverage prior knowledge (i.e., demonstration examples) to generate comments without parameter updating.
However, the effectiveness of in-context learning heavily relies on the quality and quantity of demonstration examples provided~\cite{geng2024large,liu2023pre}.

Based on the above research motivations, we propose a novel approach {\tool} ( \underline{S}mart \underline{C}ontract \underline{C}omment Generation via \underline{L}arge \underline{L}anguage \underline{M}odels), which mainly contains three phases.
In particular, we employ a customized two-phase retrieval strategy during the demonstration selection phase. This strategy allows us to retrieve the top-$k$ high-quality demonstration examples from a historical corpus, considering the semantic, syntactic, and lexical information of the code snippets.
Subsequently, in the in-context learning phase, we leverage these retrieved top-$k$ demonstrations to construct a customized prompt. By incorporating these demonstrations, we can utilize the knowledge related to smart contract comment generation within LLMs through in-context learning.
Once the prompt is constructed, we proceed to the LLMs inference phase. Here, we directly utilize the interface of LLMs, providing the customized prompt along with the target smart contract code snippet as input. The output is then generated by LLMs, representing the corresponding comment for the given code snippet.

To evaluate the effectiveness of our proposed approach {\tool}, we conduct extensive experiments on a dataset with 29,720 $\langle$method, comment$\rangle$ pairs, which were gathered from 40,933 smart contracts in a smart contract community Etherscan\footnote{\url{https://etherscan.io/}}.
We use ChatGPT\footnote{\url{https://chat.openai.com/}} as the representative LLM due to its promising performance for code intelligence tasks (such as automated program repair~\cite{xia2022less, xia2023automated,xia2023keep}, automatic code generation~\cite{dong2023self,liu2023improving}).
In our empirical study, 
we first compare {\tool} with three state-of-the-art baselines~\cite{yang2022ccgir,wang2021codet5,zhang2020retrieval} in terms of automatic performance measures.
For example, {\tool} can average improve the performance by 7.70\%, 8.14\%, 2.49\%, and 17.26\% in terms of BLEU, ROUGE-1, ROUGE-2, and ROUGE-L respectively.
Moreover, we show the effectiveness of our customized demonstration selection strategy through ablation studies.
Our ablation studies show that our used strategy can help to select higher-quality demonstration examples when compared to a set of control strategies, which were designed to evaluate the rationality of the component settings in our customized strategy.
Later, we also analyze the influence of the number of demonstrations on {\tool} and find that the performance of {\tool} is low when only a small number of demonstrations are provided. However, when more high-quality demonstrations are provided, the performance of {\tool} can be substantially improved, which can eventually outperform state-of-the-art baselines.
Since the automatic performance measures can only reflect the lexical similarity between the generated smart contract comment and the ground-truth smart contract comment, we finally conduct a human study to evaluate the quality of the generated comments. 
By following the human study methodology considered in the previous study for similar tasks~\cite{mu2022automatic,roy2021reassessing}, we find {\tool} can generate higher-quality comments than baselines in terms of similarity, naturalness, and informativeness perspectives.

Our automatic and human evaluation results show using LLMs and in-context learning is a promising direction to improve the smart contract comment quality. 
Therefore, we hope that more researchers can conduct follow-up research in this promising direction, and our proposed approach can be also customized for other software document generation tasks (such as commit message generation~\cite{liu2018neural,tao2022large}, issue title generation~\cite{chen2020stay,lin2023gen}, and pull request title generation~\cite{zhang2022automatic}).

The main contributions of our study can be summarized as follows:

\begin{itemize}
    \item \textbf{Direction.} Recently, LLMs have shown high performance in different software engineering tasks (such as program repair, code generation, and test case generation)~\cite{wang2023software,hou2023large}. 
    However, to our best knowledge, the potential of LLMs to enhance the performance of smart contract comment generation has not been thoroughly investigated in previous studies.
    In light of the promising results shown by our research, we encourage more follow-up studies to the exploration of using LLMs in this specific task.
    
    \item \textbf{Approach.} We propose a novel approach {\tool} based on the representative LLM (i.e., ChatGPT). Specifically, {\tool} uses an effective customized retrieval strategy for selecting top-$k$ high-quality demonstrations and performs in-context learning by these retrieved demonstrations. After the in-context learning, {\tool} can generate a corresponding comment for the target smart contract code snippet.
    
    \item \textbf{Study.} We conducted a comprehensive empirical study on the dataset with 29,720 $\langle$method, comment$\rangle$ pairs.
    Comparison results based on automatic performance measures and human studies show the effectiveness of {\tool}. Ablation studies also show the rationality of our customized demonstration selection strategy.
    
\end{itemize}

To encourage the follow-up studies for applying LLMs to smart contract code comment generation, we share data, code, and detailed results at our project home: 

\url{https://github.com/jun-jie-zhao/SCCLLM}.

The rest of this paper is organized as follows.
Section~\ref{sec:approach} shows the framework and details of our proposed approach {\tool}. 
Section~\ref{sec:setup} shows the empirical settings of our study, including research questions and design motivation, experimental subjects, performance measures, baselines, implementation details, and running platform. 
Section~\ref{sec:result} presents our result analysis for research questions. 
Section~\ref{sec:discussion} discusses the limitations of our study and threats to validity analysis. 
Section~\ref{sec:rw} summarizes related studies to our work and emphasizes the novelty of our study. 
Finally, Section~\ref{sec:conclusion} concludes our study and shows potential future directions.

\section{Our Proposed Approach}
\label{sec:approach}

\begin{figure*}
	\centering
	\includegraphics[width=0.95\textwidth]{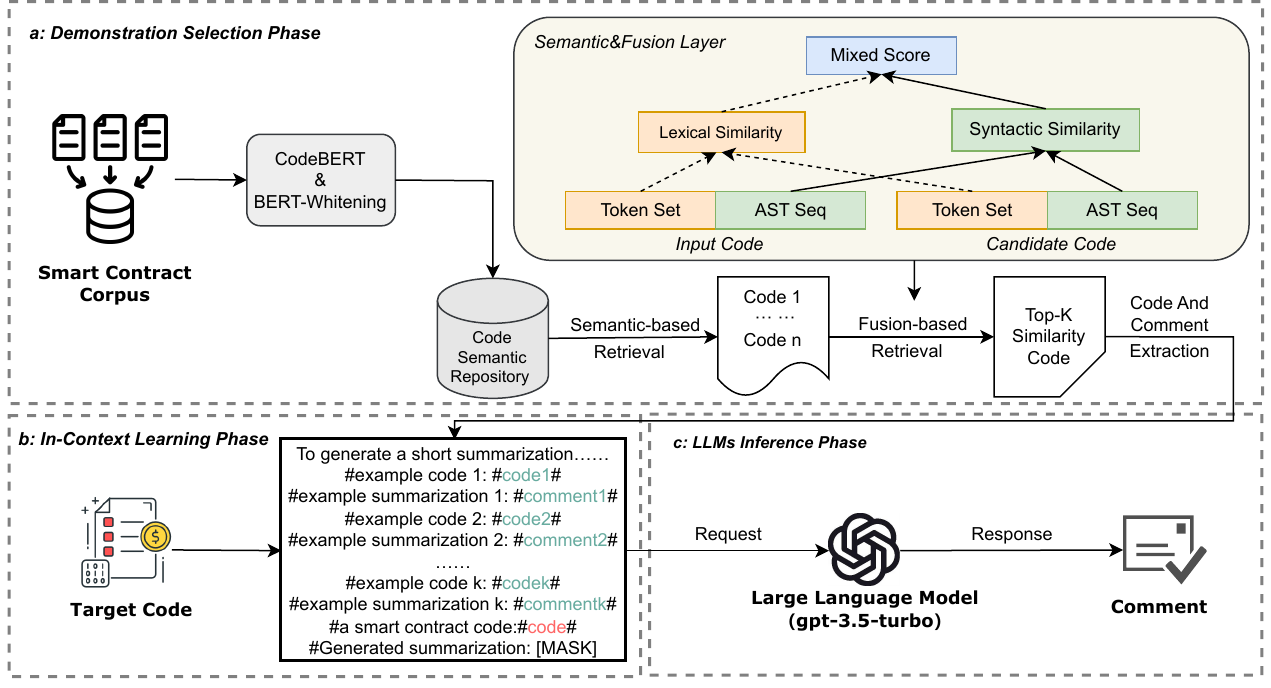}
	\caption{Framework of our proposed approach {\tool}}
	\label{fig:framework}
\end{figure*}

We show the overall framework of our proposed approach {\tool} in Figure~\ref{fig:framework}.
In this figure, we can find that {\tool} mainly contains three phases. 
Specifically, in the \textbf{demonstration selection phase}, 
{\tool} retrieves the top-$k$ smart contract code snippets from the historical corpus that are most similar to the target smart contract code snippet by considering syntax, semantics, and lexical information.  
In the \textbf{in-context learning phase}, 
{\tool} utilizes the retrieved top-$k$ smart contract code snippets and their associated comments as demonstration examples, which can be used to mine potentially domain knowledge related to smart contract code comment generation from LLMs by in-context learning. 
In the \textbf{LLMs inference phase},  the input is the target smart contract code snippet, and the output is the corresponding comment generated by the LLMs.
In the rest of this section, we show the details of these three phases.

\subsection{Demonstration Selection Phase}
\label{sec:demonstrationselection}

According to the recent survey for in-context learning~\cite{dong2022survey}, a customized demonstration example selection strategy can effectively utilize the domain knowledge hidden in the LLMs since high-quality demonstration examples that are highly related to the target can help LLMs better understand the investigated task. 
In previous study~\cite{geng2024large}, demonstration selection strategies have typically been designed using token-based and sequence-based methods. However, these methods only consider one type of code information during demonstration retrieval. To address this issue, our study aims to employ a novel retrieval strategy that integrates multiple types of code information. The primary challenge faced in our design is how to effectively fuse various code information types (i.e., semantic information, lexical information, and syntactic similarity).
Therefore, in this phase, we adopt our previously proposed information retrieval approach CCGIR~\cite{yang2022ccgir} as our demonstration selection strategy. By using our customized demonstration selection strategy, we can select top-$k$ similar smart contract code snippets from the historical repository by considering semantic, syntactic, and lexical information when given the target smart contract code snippet.
Our demonstration selection strategy can be divided into two parts: (1) the semantic-based retrieval part (i.e., the first part) and (2) the syntax and lexical-based retrieval part (i.e., the second part). 
Specifically, in the first part, we use CodeBERT~\cite{feng2020codebert} and BERT-whitening~\cite{su2021whitening} to extract semantic information from smart contract code snippets. Then, we retrieve the top-$n$ smart code snippets from the historical corpus that are most similar to the target smart contract code snippet as candidates. 
However, directly using these top-$n$ candidate smart contract code snippets in terms of only semantic information may ignore their structural information and lexical information. 
Therefore, in the second part, we further consider the syntax and lexical information of these top-$n$ candidates. We calculate their lexical and syntactic similarities and then obtain the top-$k$ code snippets as the final demonstration examples based on the weighted sum of these two similarities.
In the rest of this subsection, we show detailed information for these two parts.

\subsubsection{Semantic-based Retrieval Part}

In this part, we first split the smart contract codes in the historical corpus (i.e., the training set) according to the CamelCase naming convention to obtain input sequences $\left\{x_{i}\right\}_{i=1}^{N}$, which ${N}$ denotes the number of the smart contract code snippets in the historical corpus. 
Then, by following previous studies~\cite{liu2022codebert,yang2023exploitgen,liu2023automated}, we feed these sequences into CodeBERT~\cite{feng2020codebert} to obtain semantic vectors ${X_{i}} \in \mathbb{R}^{D}$, in which ${D}$ represents the hidden dimension. 
Later, we further process the semantic vectors using BERT-whitening~\cite{su2021whitening}, which uses a simple
linear transformation to enhance the isotropy of sentence representations, to reduce the dimensionality of the vector from ${D}$ to ${d}$, and perform a linear transformation to obtain $\left\{\tilde{X}_{i}\right\}_{i=1}^{N}$. 
The purpose of using BERT-whitening is to improve the quality and effectiveness of the embeddings generated by the BERT model. Previous studies show that this technique can help reduce the redundancy in the embeddings, result in faster training times, and better handle noisy data~\cite{yin2022efficient,zhuo2023whitenedcse}.
Finally, we calculate the semantic similarity between the two smart contract code embeddings ${\tilde{X}_{a}}$ and ${\tilde{X}_{b}}$ by the L2 distance, which can be calculated as follows.

\begin{equation}
\operatorname{semantic\_ similarity}\left(\tilde{X}_{a}, \tilde{X}_{b}\right)=\sum_{i=1}^{d}\left(\tilde{X}_{a}[i]-\tilde{X}_{b}[i]\right)^{2}
\end{equation}

\subsubsection{Syntax and Lexical-based Retrieval Part}

Based on semantic similarity, we can retrieve the top-$n$ candidate smart contract code snippets from the corpus. 
However, only considering semantic information based on CodeBERT and BERT-whitening may ignore the structural and lexical information in the smart contract code. Therefore, in this part, we further incorporate syntactic and lexical similarity. Specifically, we employ AST (Abstract Syntax Tree) sequences and code tokens to compute a mixed score, which can help to identify more similar smart contract code snippets.
The reason why we first consider semantic similarity in our two-stage demonstration selection strategy is that compared to lexical similarity or syntax similarity, the retrieval quality of semantic similarity is higher~\cite{yang2022ccgir,siow2022learning,yu2022bashexplainer}.

For two smart contract code snippets ${A}$ and ${B}$, we use the method SimSBT~\cite{yang2021comformer} to generate two sequences   ${\tilde{A}}$ and ${\tilde{B}}$. SimSBT is used to generate the sequence for each AST, which can better represent the structure of the AST.
We calculate the syntax similarity using the following formula:

\begin{equation}
\operatorname{syntactic\_ similarity(A,B)}=\frac{\operatorname{sum}(\operatorname{len}(\tilde{A}), \operatorname{len}(\tilde{B}))-\operatorname{lev}}{\operatorname{sum}(\operatorname{len}(\tilde{A}), \operatorname{len}(\tilde{B}))} 
\end{equation}
Where lev is the Levenshtein distance~\cite{yujian2007normalized} between sequences ${\tilde{A}}$ and ${\tilde{B}}$.

Lexical information mainly considers tokens in two smart contract code snippets. Since code snippets often contain many repeated tokens, 
to address this issue, we treat the code as a sequential structure and represent the tokens of two smart contract code snippets through sets. Based on the code sequence, we remove duplicate tokens to obtain two token sets $\text{set}_A$ and $\text{set}_B$. Then we calculate the lexical similarity by the Jaccard similarity.

\begin{equation}
\operatorname{lexical\_similarity}(A, B) = \frac{\mid \operatorname{set}_A \cap \operatorname{set}_B \mid}{\mid \operatorname{set}_A \cup \operatorname{set}_B \mid}
\end{equation}

Based on the top-$n$ similar smart contract code snippets retrieved in the first part, we further select the top-$k$ most similar smart contract code snippets by fusing the syntax similarity and the lexical similarity as follows:

\begin{equation}
\begin{aligned}
\operatorname { mixed\_score }(A, B) = \lambda \times \operatorname { lexical\_similarity }(A, B) \\+(1-\lambda) \times \operatorname { syntactic\_similarity }(A, B)
\end{aligned}
\end{equation} 
where $\lambda$ is a parameter that can adjust the weights between different similarities.

\subsection{In-Context Learning Phase}

In-Context Learning~\cite{dong2022survey} is a novel paradigm distinct from the fine-tuning paradigm. Fine-tuning is a resource-intensive process, particularly as the current training parameters of large language models have significantly increased, leading to promising performance in many downstream tasks~\cite{liu2023large}. 
In contrast, in-context learning is a paradigm that utilizes a small number of demonstration examples to leverage the related domain knowledge in the LLMs for new tasks.
It eliminates the need for gathering massive high-quality training data for new tasks, thereby avoiding the limitations of the fine-tuning paradigm.

\begin{figure}[htbp]
	\centering
	\includegraphics[width=0.4\textwidth]{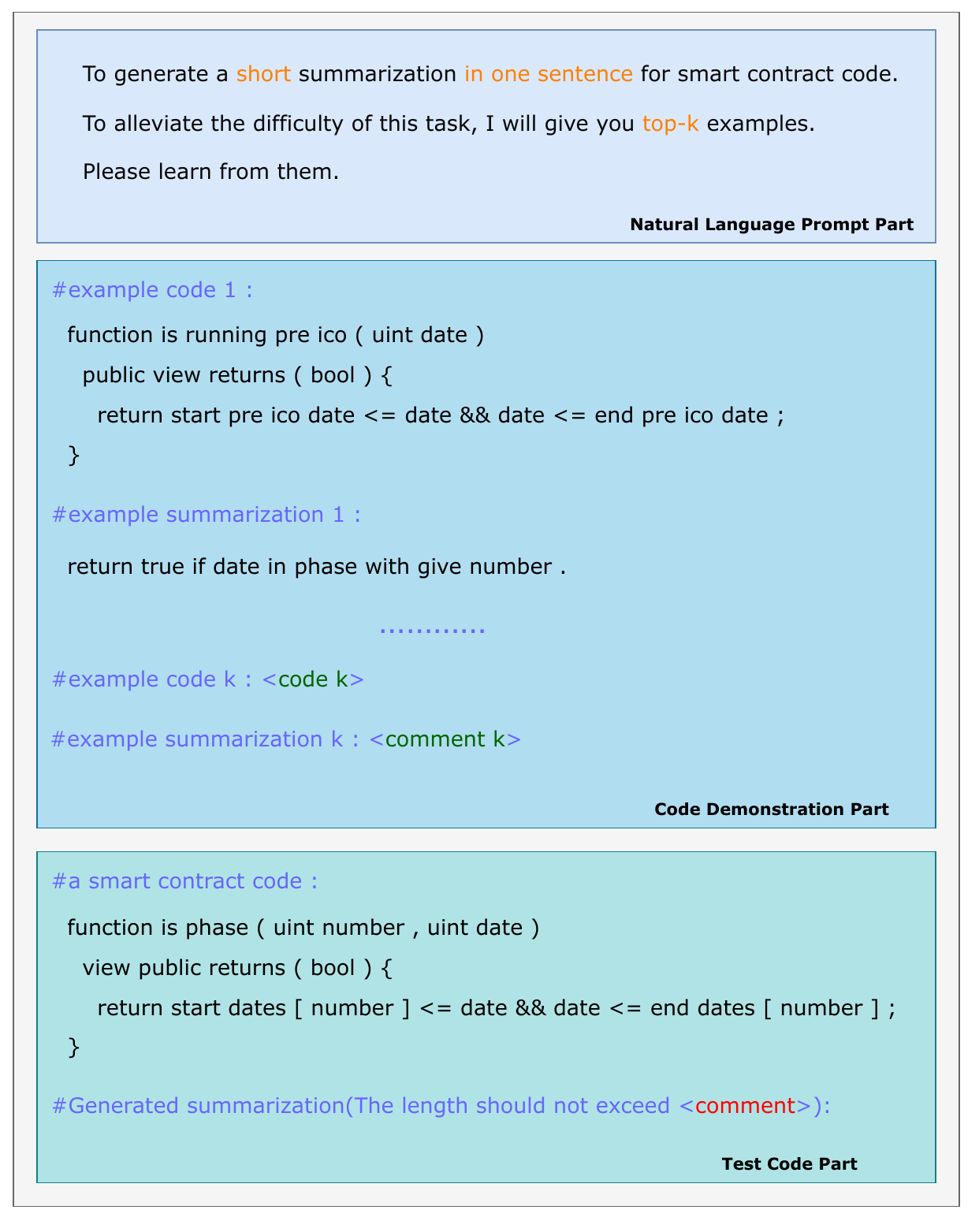}
	\caption{The prompt template used by our proposed approach {\tool}}
	\label{fig:template}
\end{figure}

However, the effectiveness of in-context learning is determined by the quality and quantity of the selected demonstration examples. 
In this phase, we utilize the top-$k$ smart contract code snippets retrieved in our customized demonstration selection phase as the demonstration examples.
Based on these examples, we can construct the prompt for in-context learning.
The prompt template used by {\tool} is shown in Figure~\ref{fig:template}.
Specifically,
the constructed prompt consists of three parts: natural language prompt part, code demonstration part, and test query part. In particular,
in the \textbf{natural language prompt part}, we first inform the LLMs to generate comments for the target smart contract code. According to the study of Sun et al. ~\cite{sun2023automatic}, a classical LLM ChatGPT, which is used in our study, often tends to generate overly lengthy comments, which may contain redundant information.
In their empirical study, they find that using  ``short", ``in one sentence", and ``no more than xx words" in the prompt can effectively limit the length of generated comments. Based on their findings, we designed our introductory content for constructing the prompt as ``To generate a short summarization in one sentence for smart contract code". To help the LLMs capture demonstration examples in in-context learning, we added `` To alleviate the difficulty of this task, we will give you top-k examples. Please learn from them" in this part. 
In the \textbf{code demonstration part}, we use ``\#" to separate comments in the natural language and smart contract code, and add the top-$k$ demonstration examples in sequence to this prompt, which can help to utilize the related domain knowledge for smart contract comment generation in the LLMs.
Finally, in the \textbf{test code part}, we input the target smart contract code and provide a prompt for generating a corresponding comment. Notice that to further prevent the LLMs from generating lengthy comments, we add the prompt ``The length should not exceed $\langle$ comment $\rangle$" after generating the comment. The purpose of this setting is to further limit the length of the generated comment. Notice in our study, we fill the comment of the retrieved code with the highest similarity into ``comment".

\subsection{LLMs Inference Phase}

For previous deep learning-based methods~\cite{yang2021multi}, the performance depends on high-quality labeled data, which was time-consuming and laborious for the data labeling process. 
With the continuous development of ChatGPT, API invocation has also become the main means of using ChatGPT for various tasks. 
Our proposed approach {\tool} does not require any model training 
and can directly generate code comments by calling the API interface. 
Specifically, we input the constructed prompt directly through the API gpt-3.5-turbo provided by OpenAI and then get the generated comment for the target smart contract code. 
The API gpt-3.5-turbo is the current mainstream version of ChatGPT, which is trained on more training data and has lower usage costs.

\section{Experimental Setup}
\label{sec:setup}

In this section, we show the details of our experimental setup, including research questions and their design motivation, experimental subject, performance measures, state-of-the-art baselines, implementation details, and running platform.

\subsection{Research Questions}

To evaluate the effectiveness of our proposed approach {\tool} and the rationality of the component setting in {\tool}, we design the following four research questions (RQs).

{\RQone}

\textbf{Motivation.} 
In this RQ, we want to investigate whether {\tool} can generate higher-quality smart contract code comments than state-of-the-art baselines.
Therefore, we select CCGIR~\cite{yang2022ccgir}, CodeT5~\cite{wang2021codet5}, and Rencos~\cite{zhang2020retrieval} as the state-of-the-art baselines.
To evaluate the quality of smart contract code comments generated by different approaches automatically, we consider BLEU~\cite{papineni2002bleu}, ROUGE-1, ROUGE-2 and ROUGE-L~\cite{lin2004rouge} as our automatic performance measures.

{\RQtwo}

\textbf{Motivation.} In previous studies~\cite{geng2024large,zhao2021calibrate,gao2021making}, the researchers found that the quality of demonstrations can have a significant impact on the effectiveness of in-context learning. Therefore, in this RQ, we want to investigate whether our customized demonstration selection strategy can help to select high-quality demonstrations, which can further improve the performance of {\tool}. 
Specifically, we compare our demonstration selection strategy with three control approaches, which can investigate the influence of demonstration selection, BERT-whitening usage, and fusion layer usage by considering syntactic similarity and lexical similarity. 

{\RQthree}

\textbf{Motivation.} In RQ2, we mainly analyze the influence of different demonstration selection strategies. In this RQ, we want to further investigate the influence of the number of demonstration examples on the effectiveness of {\tool}.

{\RQfour}

\textbf{Motivation.}
Performance measures (such as BLEU~\cite{papineni2002bleu}, ROUGE-1, ROUGE-2, and ROUGE-L~\cite{lin2004rouge}) can only evaluate the lexical similarity between the generated smart contract comments and the ground-truth comments.
However, these performance measures are inadequate in reﬂecting the real semantic differences for comments~\cite{haque2022semantic}.
Therefore, we want to conduct a human study for smart contract comment quality evaluation for different approaches in this RQ by considering similarity, naturalness, and informativeness perspectives.

\subsection{Experimental Subject}

\begin{figure}
	\centering
	\includegraphics[width=0.45\textwidth]{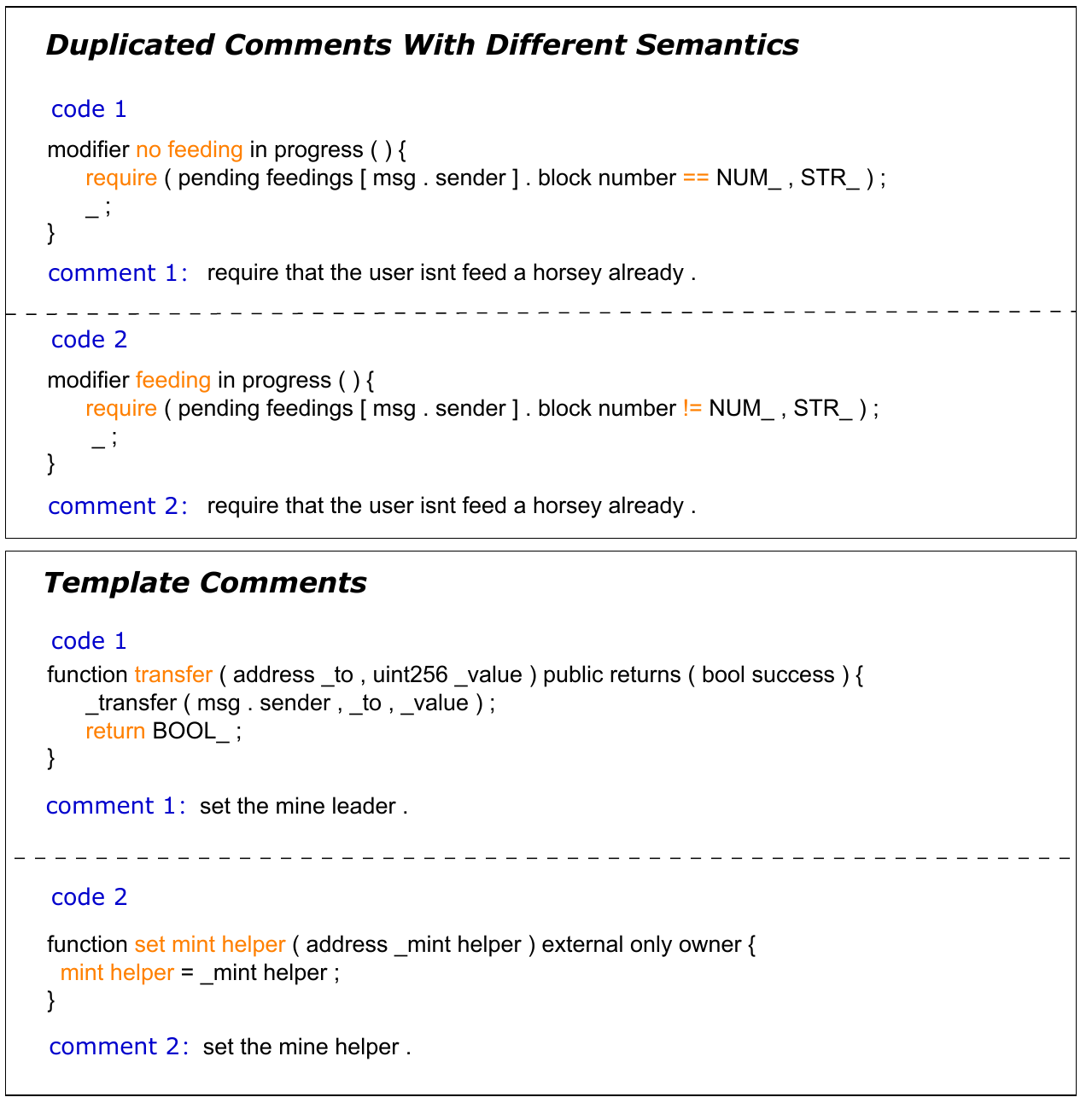}
	\caption{Examples of two problems in the dataset shared by Yang et al.~\cite{yang2021multi}.}
	\label{fig:data example}
\end{figure}

The raw data of our experimental subject was originally shared by Zhuang et al.~\cite{zhuang2021smart} for studying smart contract vulnerability detection, which was gathered from 40,932 smart contracts written in solidity on a popular and active smart contract community Etherscan.io\footnote{\url{https://etherscan.io/}}. 
Then this raw data was processed by Yang et al.~\cite{yang2021multi} for studying the smart contract comment generation. For example, they only considered normal functional methods and modifiers. They removed smart contract codes, which contain less than four words.
However, after the manual analysis in our study, we found there were still low-quality pairs in their processed dataset: 
(1) There are smart contract code snippets with different semantics but the duplicated comments, and 
(2) There are template comments, which may be automatically generated by smart contract development tools or paste copy behavior from developers. 
Figure~\ref{fig:data example} illustrates corresponding cases for these two problems. 
In the case of duplicated comments with different semantics, the two code snippets in this Figure have the same comments, but they have different semantics.
In code 1, the triggering condition for the modifier is when the block number is NULL, whereas in code 2, the triggering condition is when the block number is not NULL. 
Although these two smart contract code snippets have different semantics, their corresponding comments are the same.
In the case of template comments, these comments appear more frequently in the dataset than other comments and these comments cannot provide very clear semantics. However, the two smart contract code snippets shown in this Figure clearly have different semantics and need more clarified and concise comments.
To improve the quality of our experimental subject, 
we removed low-quality pairs with these problems as many as possible in a manual way.
Finally, we obtain 29,720 $\langle$method, comment$\rangle$ pairs in our experimental subjects. 

By following the previous experimental setting for smart contract code comment generation studies~\cite{zhuang2021smart,yang2021multi}, we split the dataset into the training set (80\%), the validation set (10\%), and the testing set (10\%).
Notice, different from baselines, our proposed approach {\tool} does not need to use the validation set.

The statistical information of our experimental subject can be found in Table~\ref{tab:dagaset}. 
After the dataset partitioning, the dataset is divided into 23,776 pairs for the training set, 2,972 pairs for the validation set, and 2,972 pairs for the testing set. 
Moreover, we also show the average (Avg.) tokens in the smart contract code snippets and comments for different sets.

\begin{table}[htbp]
\scriptsize
  \centering
  \caption{Statistical information of our experimental subject}
 \resizebox{0.45\textwidth}{!} {
    \begin{tabular}{cccc}
    \toprule
    \textbf{Statistic} & \textbf{Train} & \textbf{Validation} & \textbf{Test} \\
    \midrule
    Number & 23,776 & 2,972 & 2,972 \\
    Avg. tokens in codes & 80.54 & 80.13 & 82.27 \\
    Avg. tokens in comments & 12.05 & 11.97 & 12.1 \\
    \bottomrule
    \end{tabular}%
    }
  \label{tab:dagaset}%
\end{table}%

\subsection{Performance Measures}

To evaluate the performance of {\tool} and baselines, we consider automatic performance measures, such as BLEU, ROUGE-1, ROUGE-2, and ROUGE-L. These performance measures can effectively evaluate the lexical similarity between the generated smart contract code comments and the ground-truth comments and have been widely used in previous smart contract code comment generation studies~\cite{yang2022ccgir,yang2021multi} and similar generation tasks for software engineering (such as source code summarization~\cite{hu2018deep,yang2021comformer}, Stack Overflow title generation~\cite{gao2020generating,liu2022sotitle}, issue title generation~\cite{chen2020stay,lin2023gen}, code generation~\cite{sun2020treegen,yang2021fine,yang2022dualsc}).
We show the details of these performance measures as follows.

\begin{itemize}

    \item \textbf{BLEU.} BLEU~\cite{papineni2002bleu} (Bilingual Evaluation Understudy) is a machine translation metric that measures the text similarity between two texts by measuring the overlap of $n$-grams. In our study, we select the BLEU-4 variant ($n$-gram precision of 4) to measure the quality of the generated smart contract comments.

    \item \textbf{ROUGE-N.} ROUGE (Recall-Oriented Understudy for Gisting Evaluation)-$N$~\cite{lin2004rouge}($N$ refers to $n$-gram, with values 1, 2, 3, 4) is an automatic evaluation measure based on $n$-grams. It assesses the quality of comments by counting the number of overlapping basic units between the generated and ground-truth comments. In our study, we only select ROUGE-1 and ROUGE-2, as they are well-suited for short comments and accurately capture text similarity.

    \item \textbf{ROUGE-L.} ROUGE (Recall-Oriented Understudy for Gisting Evaluation)-L~\cite{lin2004rouge} is an evaluation measure for measuring the similarity between the ground-truth comment and the generated comment by comparing their longest common subsequence. 

\end{itemize}

For these performance measures, the value range is between 0 to 1, and the values are displayed as a percentage.
Notice the higher the value of these performance measure values, the closer the generated comment is to the ground-truth comment (i.e., the better performance of the corresponding approach).
To alleviate the internal threat due to implementation errors for these performance measures, we use nlg-eval library\footnote{\url{https://github.com/Maluuba/nlg-eval}} to compute BLEU measure and Rouge library\footnote{\url{https://github.com/pltrdy/rouge}} to compute ROUGE measures. 

\subsection{Baselines}

To evaluate whether our proposed approach {\tool} can achieve state-of-the-art performance, we consider the following three baselines related to our study.

\begin{itemize}
    \item \textbf{CCGIR}. CCGIR~\cite{yang2022ccgir} is an information retrieval technique for smart contract code comment generation that demonstrates superior performance among different information retrieval methods. It employs an information retrieval approach to retrieve the most similar smart contract code in the historical repository by considering semantic similarity, lexical similarity, and syntactic information. Finally, it reuses the comments associated with the retrieved most similar smart contract code.

    \item \textbf{CodeT5}. CodeT5~\cite{wang2021codet5} is an encoder-decoder transformer model that is pre-trained based on T5~\cite{raffel2020exploring}. Compared to other deep learning models, it exhibits better comprehension of code information and possesses stronger generation capabilities. This model employs a unified framework to seamlessly support code understanding and generation tasks, while also enabling multitask learning, thereby demonstrating excellent performance across various downstream tasks.

    \item \textbf{Rencos}. Rencos~\cite{zhang2020retrieval} is a hybrid approach for source code summarization that combines information retrieval and deep learning. Specifically, Rencos not only trains an attention-based encoder-decoder model using code snippets and comments from the training set but also incorporates two most similar code snippets retrieved based on semantic and syntactic similarities. During the encoding phase, the input code is combined with the two most similar code snippets. Finally, during the decoding phase, the comment is generated by incorporating the fused information.
    
\end{itemize}

Notice the first baseline CCGIR can be treated as the state-of-the-art baseline for smart contract code comment generation. In our previous study~\cite{yang2022ccgir}, we find CCGIR can significantly outperform the smart contract code comment generation approach MMTrans~\cite{yang2021multi}. Therefore, we do not consider MMTrans as our baseline. 
To conduct a comprehensive evaluation, we also consider a representative deep learning-based baseline (i.e., CodeT5~\cite{wang2021codet5}) and a representative hybrid baseline (i.e., Rencos~\cite{zhang2020retrieval}) for our investigated generation task.

To alleviate the internal threats, we utilize the scripts shared by these baselines~\cite{yang2022ccgir,zhang2020retrieval} and follow the hyperparameter settings suggested in their original studies. For the pre-trained model CodeT5, we implement it with Hugging Face\footnote{\url{https://huggingface.co/Salesforce/codet5-base}}.

\subsection{Implementation Details}

In our experiments, the detailed parameter settings in our demonstration selection phase can be found in Table~\ref{tab:value}. These values are configured based on the suggestions from previous studies~\cite{yang2022ccgir,su2021whitening} and the optimization of our experimental results.

\begin{table}[htbp]
\tiny
  \centering
  \caption{The configuration of Hyper-parameters in our demonstration selection phase}
 \resizebox{0.45\textwidth}{!} {
    \begin{tabular}{cc}
    \toprule
    \textbf{Hyper-parameter} & \textbf{Value} \\
    \midrule
    Maximum input length of code snippet ${x_{a}}$ & 256 \\
    Dimension ${D}$ before BERT-whitening & 768 \\
    Dimension ${d}$ after BERT-whitening & 256 \\
   { mixed\_score } coefficient $\lambda$ & 0.7 \\
    Number of top-$n$ candidates & 10 \\
    \bottomrule
    \end{tabular}%
    }
  \label{tab:value}%
\end{table}%

In the LLMs inference phase, we only select five demonstration examples to construct the prompt for the in-context learning. The reason is the prompt size of the LLMs is limited, and the escalation in experimental expenses is associated with an excess of demonstrations.
For our experiments, we utilize the API interface version gpt-3.5-turbo, which has demonstrated good performance in various downstream tasks~\cite{thakur2023verigen,eliseeva2023commit}.
To guarantee a fair comparison, we also optimize the parameters of the baseline methods to achieve optimal performance.

\subsection{Running Platform}

We run all the experiments on a computer (CPU 3.50GHz) with a GeForce RTX4090 GPU (24GB graphic memory). The running operating system is Windows 10.

\section{Result Analysis}
\label{sec:result}

\subsection{RQ1: Comparison with baselines via automatic evaluation}

\textbf{Method.} To show the effectiveness of our proposed approach {\tool} in smart contract comment generation, we select the information retrieval approach CCGIR~\cite{yang2022ccgir}, the deep learning approach CodeT5~\cite{wang2021codet5}, and the hybrid approach Rencos~\cite{zhang2020retrieval} as baselines. 
Notice in this RQ, for {\tool}, we utilize the top-5 most similar code snippets as our demonstrations for in-context learning.

\begin{table}[htbp]
  \centering
  \caption{Comparison results between {\tool} and baselines in terms of four performance measures}
    \begin{tabular}{ccccc}
    \toprule
    \textbf{Approach} & \textbf{BLEU} & \textbf{Rouge-1} & \textbf{Rouge-2} & \textbf{Rouge-L} \\
    \midrule
    CCGIR & 31.21 & 32.94 & 17.33 & 24.05 \\
    CodeT5 & 30.93 & 32.63 & 16.89 & 23.69 \\
    Rencos & 30.92 & 32.58 & 16.78 & 23.62 \\
    {\tool} & \textbf{33.41} & \textbf{35.38} & \textbf{17.42} & \textbf{27.89} \\
    \bottomrule
    \end{tabular}%
  \label{tab:baselinesresults}%
\end{table}%

\textbf{Result.} Table~\ref{tab:baselinesresults} presents the comparison results between {\tool} and the baselines for smart contract comment generation. 
In this table, we emphasize the best performance for different performance measures in bold.
Specifically, {\tool} can achieve the performance of 33.41\%, 35.38\%, 17.42\%, and 27.89\% in terms of BLEU, ROUGE-1, ROUGE-2, and ROUGE-L performance measures.
Compared to the baselines, the performance of {\tool}  can be improved on average by 7.70\%, 8.14\%, 2.49\%, and 17.26\% in terms of four performance measures.
Then to check whether the performance difference between {\tool} and baselines is significant, we conduct Wilcoxon signed-rank tests~\cite{wilcoxon1992individual} at the confidence level of 95\%. Our $p$-value is smaller than 0.05, which means the performance improvement of {\tool} compared to baselines is significant.

When comparing two baselines CodeT5 and Rencos, we find they achieve similar results in our study. Although the experimental results of Rencos~\cite{zhang2020retrieval} indicate that the hybrid approach can outperform the deep learning approach. However, we consider CodeT5 as the deep learning-based baseline in our study, which is a state-of-the-art code pre-trained model, while the deep learning model part of Rencos only considers the attentional encoder-decoder model, which is trained from scratch.
It is worth noting that the information retrieval approach CCGIR still outperforms the deep learning approach CodeT5 and the hybrid approach Rencos in terms of all the performance measures in our study.
This finding is consistent with our previous research findings~\cite{yang2022ccgir} due to the extensive code reuse (i.e., code clone) during smart contract development. 
However, our proposed approach {\tool} can outperform CCGIR, which shows applying LLMs to smart contract generation is a valuable direction and worth paying attention to.
Here, we find all the approaches achieve low performance for the ROUGE-2 measure. The possible reason is that ROUGE-2 is mainly focused on 2-gram overlap when compared to ROUGE-1 and most generation-based approaches tend to generate lengthy code comments. 

\begin{figure}
	\centering
	\includegraphics[width=0.45\textwidth]{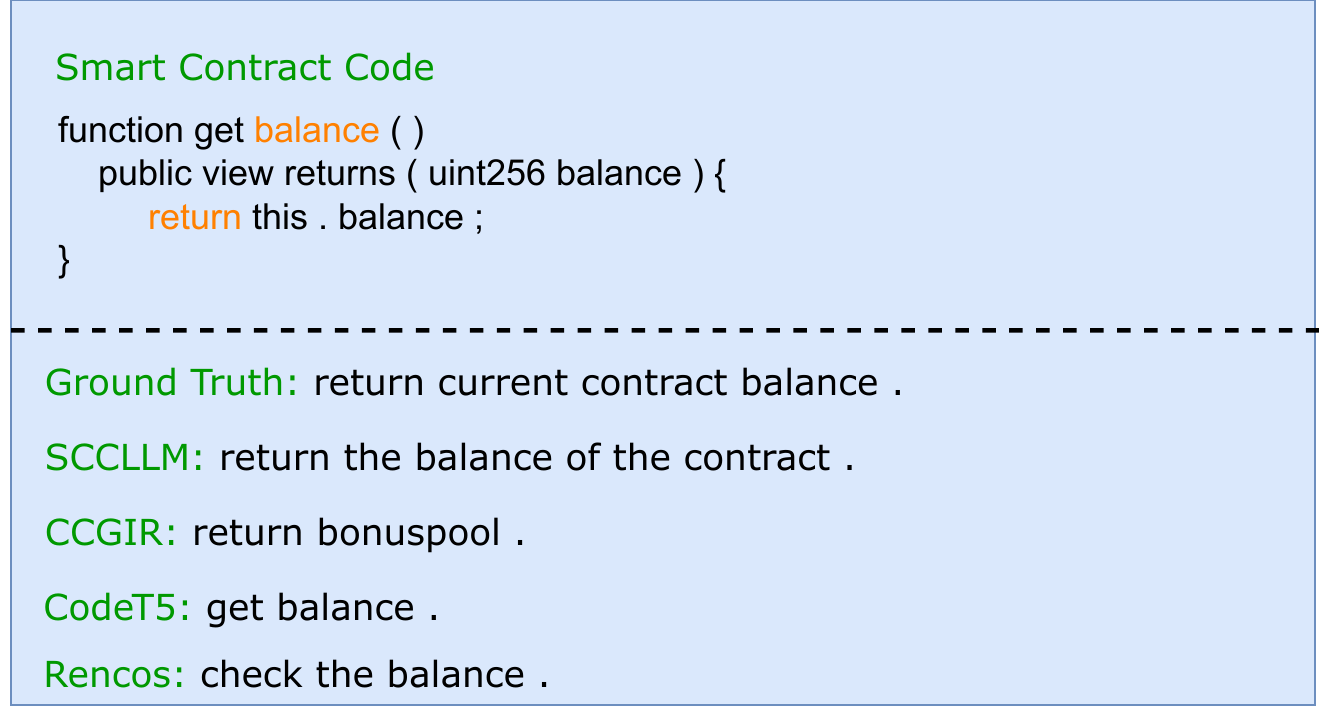}
    \includegraphics[width=0.45\textwidth]{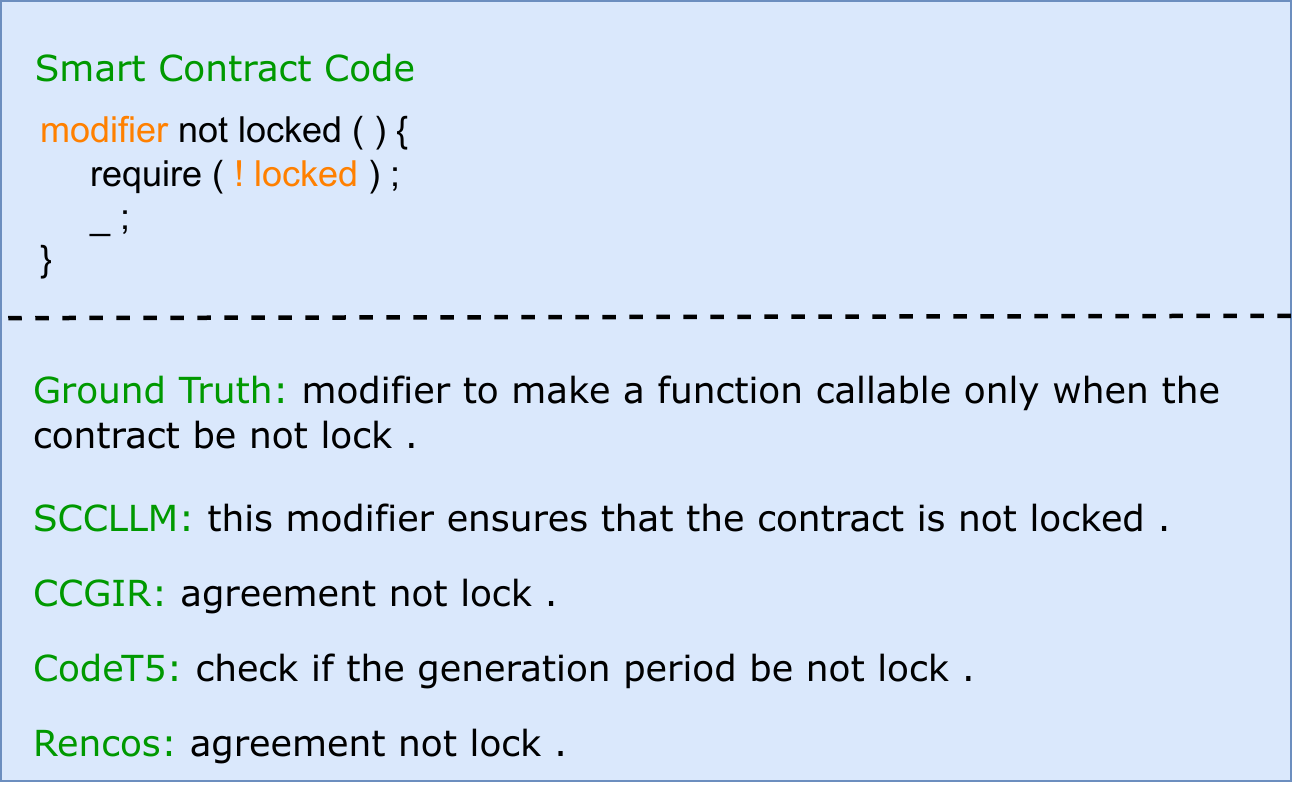}
	\caption{Comments generated by {\tool} and baselines for two cases}
	\label{fig:cases}
\end{figure}

Finally, we use two cases in Figure~\ref{fig:cases} to show the effectiveness of our proposed approach.
In the first case, CodeT5 and Rencos can learn the keyword ``balance" in the target smart contract code. However, these two baselines fail to understand the specific meaning of the target code. 
CCGIR cannot retrieve sufficiently similar code, so its reused comment is independent of the semantics of the target code.
While the comment generated by {\tool} conveys an identical semantic when compared with the ground-truth comment. 
In the second case, the comments generated by {\tool} as well as the ground-truth comment both contain the key phrases ``modifier" and ``not locked".
This indicates that {\tool} can effectively understand the target smart contract code and generate the corresponding comment by LLMs and in-context learning.
However, the comments generated by three baselines can only convey the basic notion of ``not lock", resulting in low-quality comments. 
Based on these two cases, we find that baselines struggle to learn the knowledge of the target smart contract code from the limited corpus, but {\tool} can leverage the related knowledge of LLMs to generate better comments for smart contract code snippets.

\begin{tcolorbox}[boxrule=1pt,boxsep=1pt,left=2pt,right=2pt,top=2pt,bottom=2pt,title=Answer to RQ1]
By combining LLMs with in-context learning, {\tool} outperforms all baselines in terms of four automatic performance measures.
For example, {\tool} can outperform the baselines by at least 7.70\% in terms of BLEU.
\end{tcolorbox}

\subsection{RQ2: Ablation study on Demonstration Selection Strategy}

\textbf{Method.} In this RQ, we want to show the effectiveness of our customized demonstration selection strategy in selecting high-quality demonstrations for in-context learning.
As introduced in Section~\ref{sec:demonstrationselection}, our customized demonstration selection strategy includes two major components. The first component utilizes CodeBERT and BERT-Whitening for semantic vector extraction, capturing code semantic information. The second component integrates AST sequences with code tokens in the fusion layer, combining syntax and lexical information to identify more similar smart contract code snippets.
Based on the component settings of our customized demonstration selection strategy, we design the following three control strategies.

\begin{itemize}
    \item \textbf{The first control strategy}. To show the significance of our customized demonstration selection strategy for {\tool}, we randomly select $k$ smart contract code snippets from the historical repository and use ``w/o DSS" to denote this control strategy. 
    
    \item \textbf{The second control strategy}. In this control strategy, we aim to investigate whether using the BERT-whitening~\cite{su2021whitening} can help to select demonstrations with higher quality and use `` w/o BERT-whitening" to denote this control strategy. Specifically, we remove the BERT-whitening operation in our demonstration selection strategy and directly utilize CodeBERT to generate code embeddings without any optimization or dimensionality reduction.
    
    \item \textbf{The third control strategy}, In this control strategy, we aim to investigate whether further considering the syntactic similarity and lexical similarity can help to select demonstrations with higher quality and use `` w/o (Lexical \& Syntactic Similarity)" to denote this control strategy. Specifically, we remove the fusion layer based on lexical and syntactic similarity from our demonstration selection strategy, which can retrieve the top $k$ most similar examples by only considering code semantic information.
\end{itemize}

Due to the high economic cost of calling ChatGPT to perform ablation experiments when considering all the smart contract code snippets in the test set, we employ a sampling method~\cite{singh2013elements} that randomly selects samples from the test set. The formula for the sampled number can be computed as follows:

\begin{equation}
M I N=\frac{n_{0}}{1+\frac{n_{0}-1}{\text { size }}}
\end{equation}
Where ${n_{0}}$ represents the confidence level, and ${ size }$ is the size of the test set. There is an error range 
$\left(\frac{Z^{2} \times 0.25}{e^{2}}\right)$
for ${n_{0}}$, where ${e}$ is the hyper-parameter and ${Z}$ is the confidence level score. In RQ2, we chose the smallest sample based on a confidence level of 95\% and ${e}$=0.05. In the end, we needed to randomly select 340 samples from the test set according to this formula.

\textbf{Result.} We show our ablation study results in Table~\ref{tab:irresults}. 
From this table, we observe that when randomly providing demonstrations without using the customized demonstration selection strategy, the quality of comments directly generated by LLMs is low, with a performance drop of 39.99\%, 45.17\%, 80.45\%, and 49.65\%, respectively in terms of four performance measures. This highlights 
using high-quality demonstrations in in-context learning can help to fully utilize the related domain knowledge in LLMs.

For the second control strategy, we find the performance of {\tool} decreases when the code embeddings are not optimized by BERT-whitening, with a performance drop of 10.05\%, 10.52\%, 17.57\%, and 10.97\%, respectively. This demonstrates the effectiveness of using BERT-whitening in improving the retrieval ability of our demonstration selection strategy in selecting higher-quality demonstration examples.

For the third control strategy, we find the performance of {\tool} decreases when removing the syntax and lexical-based retrieval part in our demonstration selection strategy,  with a performance drop of  
11.79\%, 13.36\%, 21.98\%, and 13.29\%, respectively. Therefore, further considering the lexical information and the syntax information can help to select higher-quality demonstration examples, which eventually improves the performance of {\tool}.

\begin{table*}[htbp]
\centering
\caption{Ablation results of {\tool} with different demonstration selection strategies}   
    \begin{tabular}{ccccc}
    \toprule
    \textbf{Demonstration Selection Strategy} & \textbf{BLEU} & \textbf{GOUGE-1} & \textbf{GOUGE-2} & \textbf{GOUGE-L} \\
    \midrule
    w/o DSS & 19.04 & 17.15 & 3.06 & 13.49 \\
    w/o BERT-whitening & 28.54 & 27.99 & 12.90 & 23.85 \\
    w/o (Lexical \& Syntactic Similarity) & 27.99 & 27.10 & 12.21 & 23.23 \\
    Our Customized Strategy & \textbf{31.73} & \textbf{31.28} & \textbf{15.65} & \textbf{26.79} \\
    \bottomrule
    \end{tabular}%
\label{tab:irresults}
\end{table*}

\begin{tcolorbox}[boxrule=1pt,boxsep=1pt,left=2pt,right=2pt,top=2pt,bottom=2pt,title=Answer to RQ2]
By using our customized demonstration selection strategy, {\tool} can help to select higher-quality demonstrations for in-context learning, which can finally improve the performance of {\tool}.
\end{tcolorbox}

\subsection{RQ3: Performance influence on the number of demonstrations}
\label{sec:resultRQ3}

\textbf{Method.} In this RQ, we want to examine the performance influence of the number of demonstrations for {\tool}. 
Specifically, we conduct experiments with different settings (i.e., zero-shot learning, one-shot learning, and few-shot learning), and compare the performance of different settings with three baselines (i.e., CCGIR~\cite{yang2022ccgir}, CodeT5~\cite{wang2021codet5} and Rencos~\cite{zhang2020retrieval}). We use ``zero-shot learning" to refer to the setting where no demonstration is provided for {\tool}. ``one-shot learning" refers to the setting where only a single demonstration is provided for {\tool}, while ``few-shot learning" refers to settings where a few demonstrations are provided for {\tool}.

Due to the maximum prompt size limitation for the API gpt-3.5-turbo, we use at most five demonstrations to construct the prompt in our study. In this experiment, we investigate the performance influence of different demonstration numbers by using all the smart contract code snippets in the test set.
To guarantee a fair comparison, we use the same prompt template shown in Figure~\ref{fig:template}, but the difference is that the number of demonstrations differs.

\textbf{ Result.} We show the performance of {\tool} with the different number of demonstrations in Table~\ref{tab:demonstrations}. 
From this table, we find that when using zero-shot learning and one-shot learning settings, 
the performance of {\tool} is very low in terms of four performance measures.
For example, if using the zero-shot learning setting, the performance of {\tool} decreases 27.48\%, 49.71\%, 83.13\%, and 46.64\%, respectively, when compared to the fine-tuning approach CodeT5. 
Our finding indicates that {\tool} cannot effectively utilize the related domain knowledge for smart contract comment generation task in the LLMs if we at most give one demonstration for in-context learning. In this table, we find that using the zero-shot learning setting can outperform using the one-shot learning setting.  The potential reason is that for zero-shot learning, we do not provide any demonstration for the prompt and simply set the length of the generated comment should not exceed 15 words. Therefore, shorter comments may result in a higher performance of {\tool} when using zero-shot learning.

When using few-shot learning, we observe a significant performance improvement for {\tool}. Specifically, when providing three demonstration examples for in-context learning, the performance of {\tool} can achieve similar performance with CodeT5.
When providing five demonstration examples for in-context learning, {\tool} can eventually outperform all the baselines.

\begin{table}[htbp]
\setlength{\tabcolsep}{1.5mm}
\centering
\caption{The performance influence of different demonstration numbers for {\tool}}   
\begin{tabular}{ccccc}
    \toprule
    \textbf{Approach} & \textbf{BLEU} & \textbf{Rouge-1} & \textbf{Rouge-2} & \textbf{Rouge-L} \\
    \midrule
    CCGIR  & 31.21 & 32.94 & 17.33 & 24.05 \\
    CodeT5 & 30.93 & 32.63 & 16.89 & 23.69 \\
    Rencos & 30.92 & 32.58 & 16.78 & 23.62 \\
    \midrule
    with zero-shot & 22.43 & 16.41  & 2.85 & 12.64 \\
    with one-shot & 18.42 & 16.09  & 3.10 & 12.47 \\
    with 3-shot & 29.05 & 27.21 & 11.60 & 23.19 \\
    with 5-shot & \textbf{33.41} & \textbf{35.38} & \textbf{17.42} & \textbf{27.89} \\
    \bottomrule
    \end{tabular}%
\label{tab:demonstrations}
\end{table}

\begin{tcolorbox}[boxrule=1pt,boxsep=1pt,left=2pt,right=2pt,top=2pt,bottom=2pt,title=Answer to RQ3]
When providing five high-quality demonstration examples for in-context learning, {\tool} can more effectively utilize the related knowledge for smart contract comment generation and achieve better performance than baselines.
\end{tcolorbox}

\subsection{RQ4: Comparison with baselines via human study}

\textbf{Method.} Based on the findings of the recent study by Sun et al.~\cite{sun2023automatic}, the current automatic performance evaluation measures cannot be used to effectively assess the quality of comments generated by LLMs. 
To alleviate this construct threat, we perform a human study to further assess the effectiveness of our proposed approach {\tool}. 
In our human study, we mainly follow the methodology used by previous studies for source code summarization~\cite{mu2022automatic,roy2021reassessing}.
Specifically, we recruit five participants, who have at least three years of experience in smart contract development and maintenance.
These participants are senior researchers or developers, who are not co-authors of our study.
We employed the same sampling method used in RQ2 and finally randomly selected 340 smart contract code snippets in the test set.
For each smart contract code, we show the participants the ground-truth comments and the comments generated by four different approaches.
To guarantee a fair comparison, the participants do not know which approach generates the comment.
Later, we allowed the participants to use the Internet to facilitate understanding the target smart contract codes if there exist concepts or codes that they are unfamiliar with. 
Before they participated in our human study, we ensured that all involved participants received a comprehensive tutorial. This process was designed to familiarize each participant with both the task expectations and the evaluation measures at hand. This training aimed to mitigate the risk of inherent biases and inconsistency in their assessments.
Finally, we require each participant to evaluate only 20 smart contract codes in a half-day to avoid biases caused by fatigue.
Each participant is asked to rate the generated four comments from three perspectives.

\begin{itemize}
     \item \textbf{Similarity.} This perspective reflects the similarity between the generated smart code comments and the ground-truth comments.
    \item \textbf{Naturalness.}  This perspective reﬂects the ﬂuency of generated smart contract comments from grammar.
    \item \textbf{Informativeness.} This perspective reﬂects the information richness of the generated smart contract comments.
\end{itemize}

\begin{figure}
	\centering
	\includegraphics[width=0.5\textwidth]{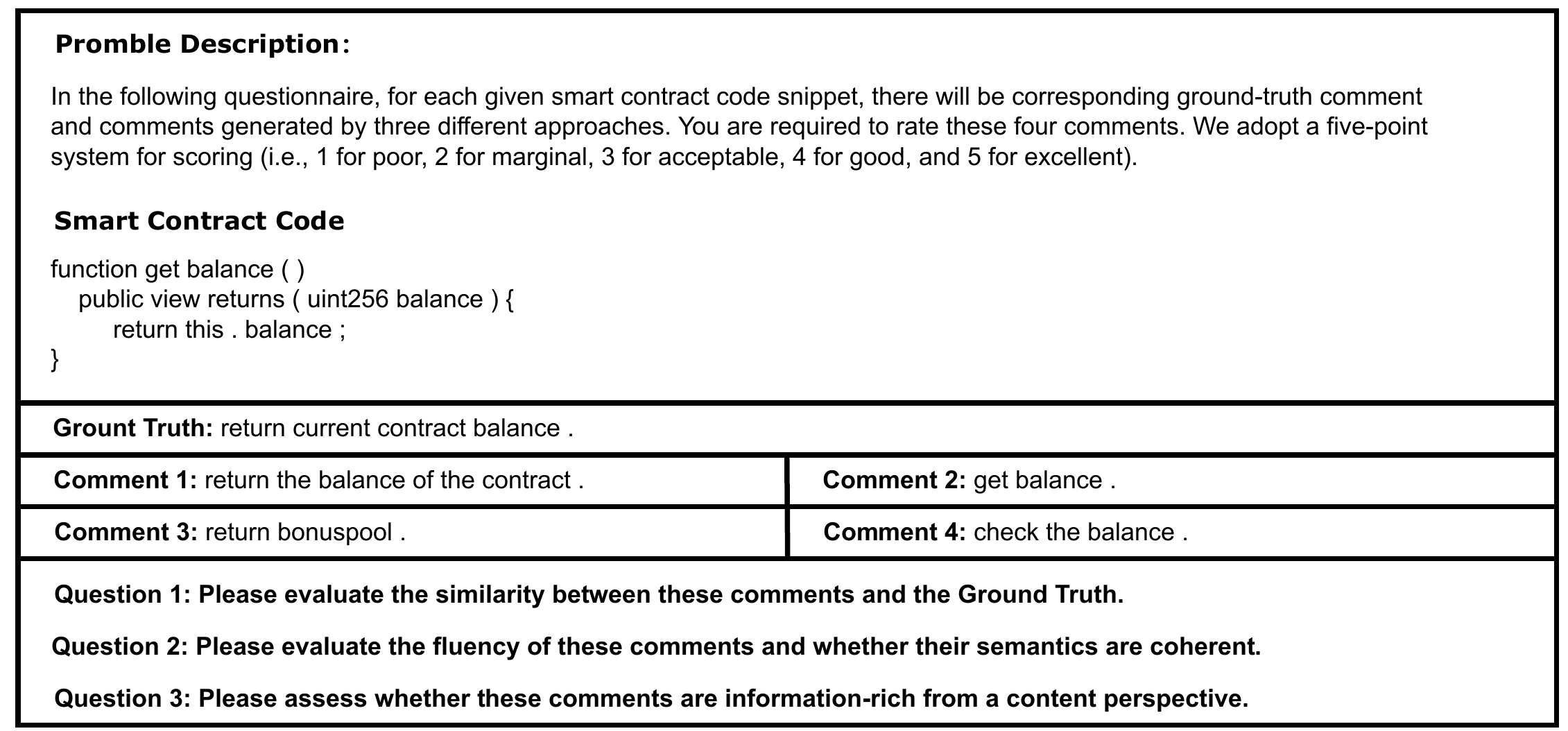}
	\caption{A questionnaire used in our human study}
	\label{fig:questionnaire}
\end{figure}

We use a five-point system for scoring (i.e.,  1 for poor, 2 for marginal, 3 for acceptable, 4 for good, and 5 for excellent). We show a questionnaire used in our human study in Figure~\ref{fig:questionnaire}.

\begin{figure}
	\centering
	\includegraphics[width=0.5\textwidth]{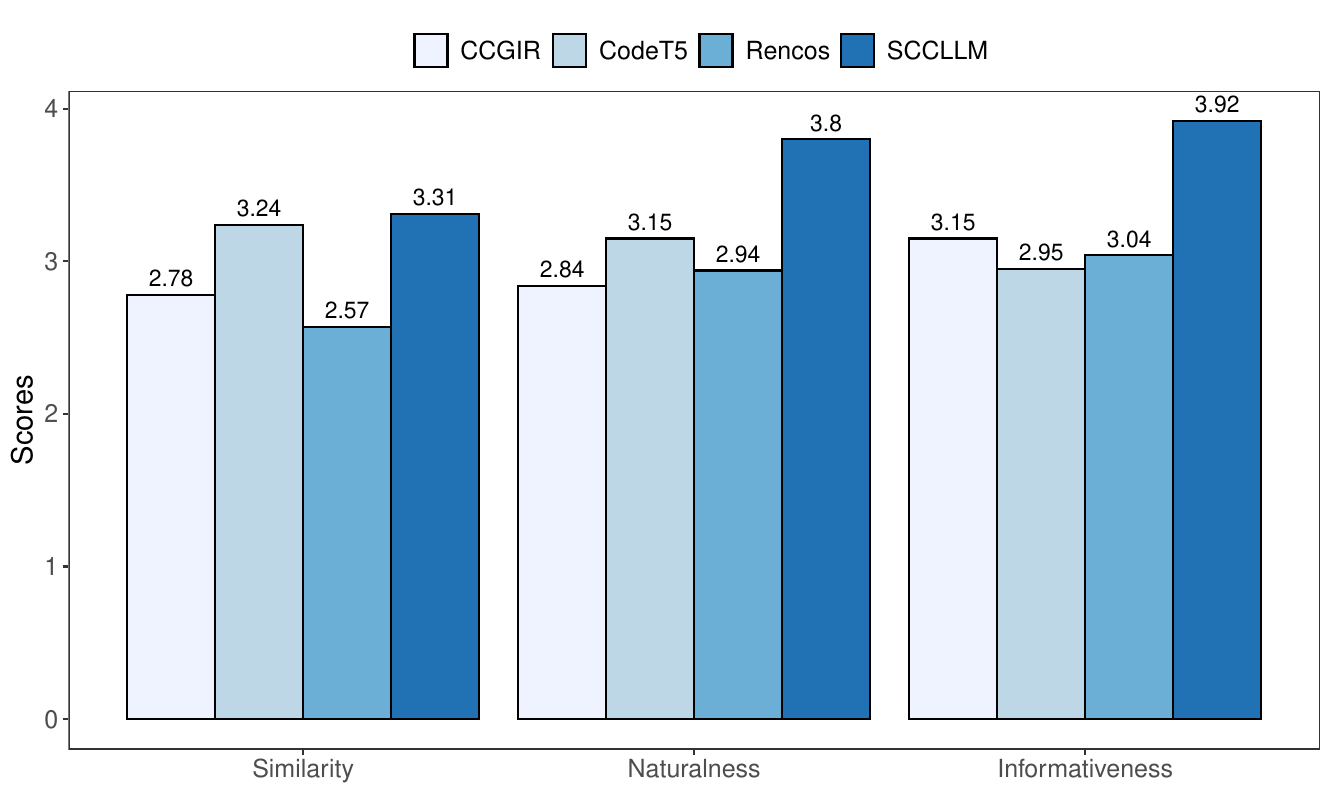}
	\caption{Our human study results between {\tool} and baselines}
	\label{fig:humanstudy}
\end{figure}

\textbf{Result.} We show our human study results in Figure~\ref{fig:humanstudy}. 
Specifically, {\tool} can achieve  3.31, 3.80, and 3.92 in terms of Similarity, Naturalness, and Informativeness.
In terms of similarity, due to the richer semantics in the smart contract comments generated by {\tool}, they exhibit a higher semantic similarity with the ground-truth comments. The result of 3.31 implies that the smart contract comments generated by {\tool} have higher quality than baselines.  In terms of naturalness and informativeness, {\tool} still performs better than all baselines, demonstrating that most of the smart contract comments generated by {\tool} are easy to understand and read, and excel in semantic comprehension. These more readable comments will help smart contract developers understand the code better and increase development efficiency.

\begin{tcolorbox}[boxrule=1pt,boxsep=1pt,left=2pt,right=2pt,top=2pt,bottom=2pt,title=Answer to RQ4]
Our human study results show that {\tool} can outperform baselines when evaluating the quality of the smart contract comments from similarity, naturalness, and informativeness perspectives.
\end{tcolorbox} 

\section{Discussions}
\label{sec:discussion}

\subsection{Limitations of {\tool}}

Though {\tool} shows competitive performance when compared to baselines for automatic evaluation and human evaluation, we also find that {\tool} may generate low-quality comments.
In this subsection, we randomly select 50 cases of this type and analyze these comments manually.
Finally, we identify the three challenging types of smart contract comment generation for {\tool}.

\begin{figure}
	\centering
	\includegraphics[width=0.5\textwidth]{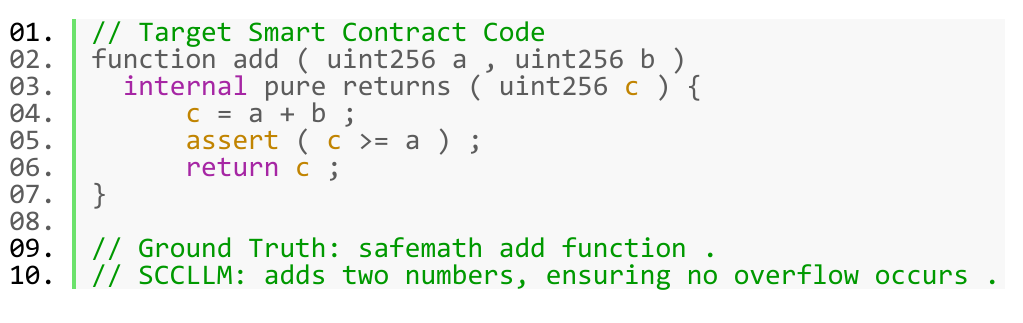}
	\caption{Case with semantic similarity but low lexical similarity.}
	\label{fig:discussion 1}
\end{figure}

\textbf{The first challenge type} is generating comments with semantic similarity but low lexical similarity. In Figure~\ref{fig:discussion 1}, we present a case to show this challenge type. 
In this case, the ground-truth comment indicates that the smart contract code snippet employs the ``add" function from the ``safemath" library, which in the Solidity language is used for secure algorithms to prevent data overflow. While the comment generated by {\tool} directly explains this code, indicating its purpose of preventing overflow. We observe that both the ground-truth comment and the comment generated by {\tool} can accurately convey the functionality of this code, but differ in the level of expertise: the ground-truth comment being more professional and the comment generated by {\tool} being straightforward and somewhat redundant.
In this case, the low lexical similarity can result in a low score in automatic evaluation. Therefore, designing new performance measures based on comment semantic similarity can alleviate this challenge type.

\begin{figure}
	\centering
	\includegraphics[width=0.5\textwidth]{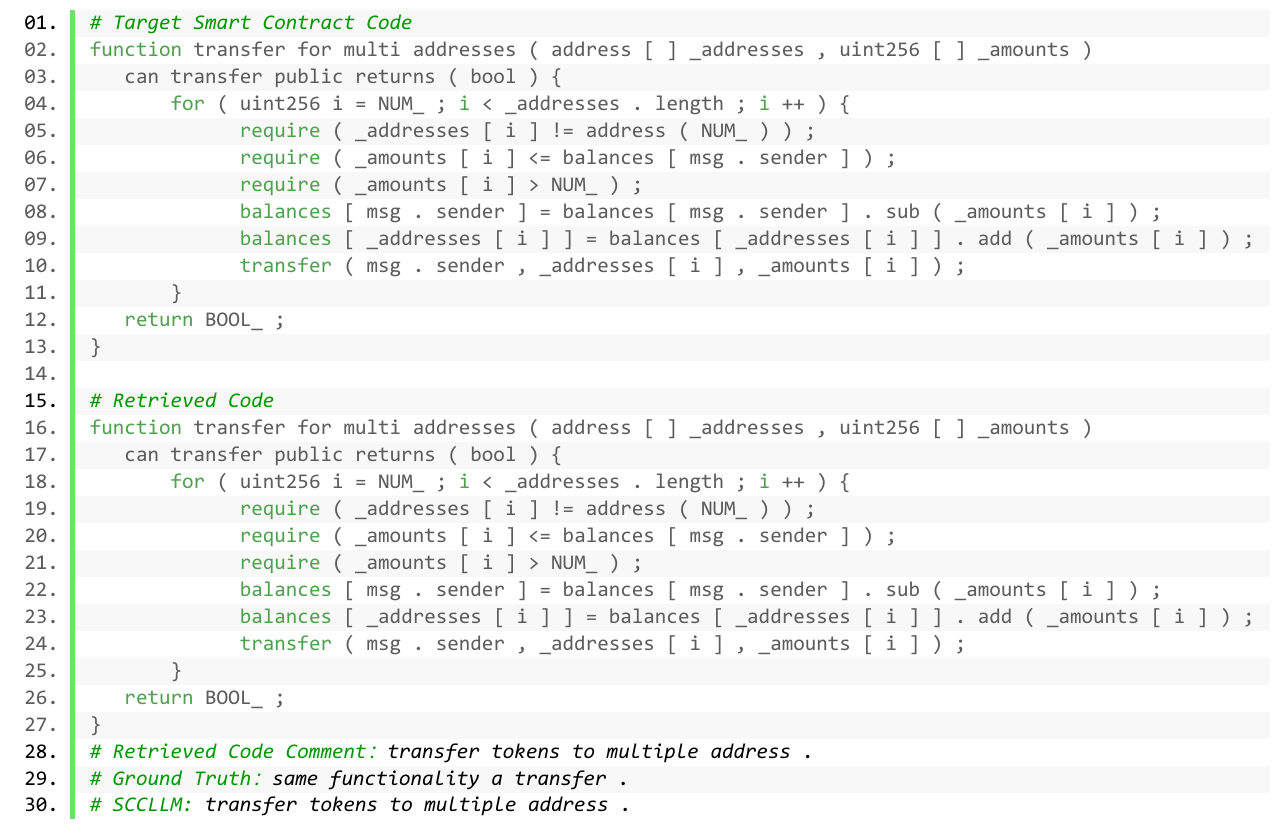}
	\caption{Case where the target smart contract code is not fully understood, and the comments from the provided demonstrations are directly reused.}
	\label{fig:discussion 2}
\end{figure}

\textbf{The second challenge type} is failing to fully comprehend the target smart contract code and directly reusing the comments from similar demonstrations. 
In our approach {\tool}, we provide high-quality demonstrations for in-context learning. However, there are cases in the provided demonstrations where the code is the same but the comments differ, leading {\tool} to directly reuse the comments.
In Figure~\ref{fig:discussion 2}, we present a case to show this challenge type.
We find that the target code and retrieved code are almost same, but correspond to different correct comments. {\tool} reuses the demonstration comments directly after learning the demonstration. Although the reused comments are semantically correct, they will result in a lower score in terms of automatic evaluation measures and will be treated as low-quality comments.

\begin{figure}
	\centering
	\includegraphics[width=0.5\textwidth]{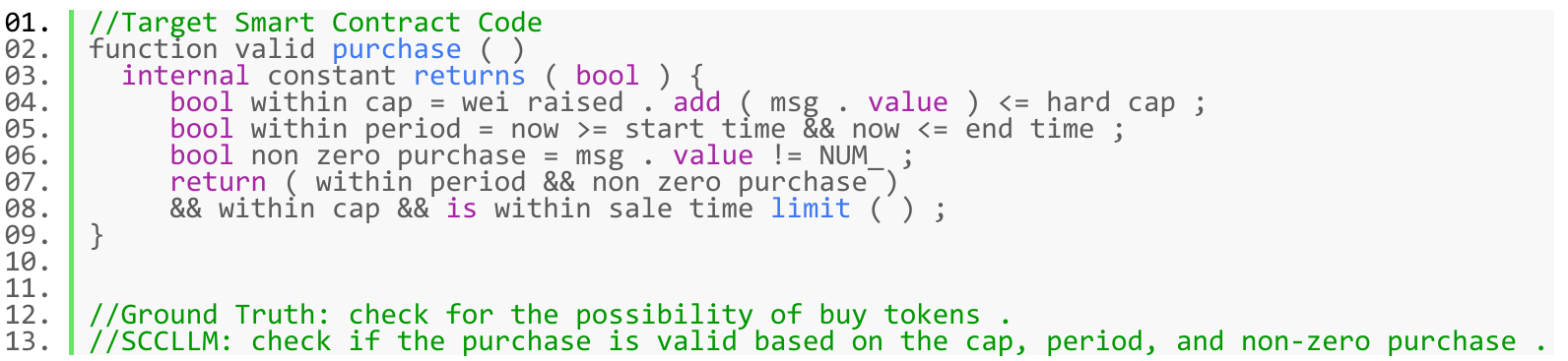}
	\caption{Case where the purposes/functionality of the smart contract code is not effectively comprehended.}
	\label{fig:discussion 3}
\end{figure}

\textbf{The third challenge type} is difficulties in effectively understanding code purposes/functionality for some smart contract code snippets.
In Figure~\ref{fig:discussion 3}, we present a case to show this challenge type.
The ground-truth comment shows the purpose of this smart contract code (i.e., ``check for the possibility of buy tokens"). However, {\tool} can only capture the words (such as ``cap", ``period", and ``non-zero purchase") in the target smart contract code, which can only provide shallow information for this code.

\subsection{Threats to Validity}

In this subsection, we mainly discuss the potential threats to our empirical findings.

\textbf{Internal threats.}
The first internal threat is the potential implementation faults in {\tool}.
To alleviate this threat, we perform code inspection in our implemented code, especially the demonstration selection strategy part.
The second internal threat is the number of demonstrations used in our study.
To alleviate this threat, we analyze the performance influence of demonstration number for {\tool} in Section~\ref{sec:resultRQ3} and find that only using five demonstrations can achieve promising performance and outperform baselines. However, improving the number of demonstrations may further improve the performance of {\tool} but at the cost of a higher escalation in experimental expenses.

\textbf{External threats.}
The first external threat is the quality of the experimental subject. To alleviate this threat, we use the gathered $\langle$method, comment$\rangle$ smart contract pairs shared by Yang~\cite{yang2021multi}.
In their study, they performed a set of processing to filter the low-quality pairs from the raw data provided by Zhuang et al.~\cite{zhuang2021smart}.
After the manual analysis in our study, we also found there still exists some low-quality pairs, which have duplicated comments but with different semantics or have template comments.
We also identify and remove these kinds of pairs, which can further improve the experimental subject quality.
The second external threat is the customized demonstration selection strategy used in {\tool}. In our study, we designed a set of experiments to verify the rationality of the component setting in this strategy. Moreover, comparison results with baselines also show the effectiveness of this strategy for retrieving high-quality demonstrations for in-context learning.

\textbf{Conclusion threats.}
The conclusion threat is related to evaluation bias in our human study. To alleviate this threat, we first invite participants who are familiar with smart contract development. Second, we provided a tutorial before our human study, which ensured that all of the participants could understand our protocol. Finally, we follow the methodology used by previous studies for a similar task (i.e., source code summarization)~\cite{mu2022automatic,roy2021reassessing} to guarantee the quality of our human study.

\textbf{Construct threats.}
The construct threat is related to the performance measures. 
To alleviate this threat, we consider four performance measures, which have been widely used in previous similar tasks, such as source code summarization~\cite{gong2022source,son2022boosting,roy2021reassessing}. 
Since the automatic measures can only evaluate the lexical similarity between the generated smart contract comments and the ground-truth comments, we further conducted a human study to evaluate the quality of the generated smart contract comments by considering similarity, naturalness, and informativeness.

\section{Related Work}
\label{sec:rw}
In this section, we present the relevant research on smart contract comment generation and recent advances in applying LLMs to Software Engineering tasks.

\subsection{Smart Contract Code Comment Generation}

To the best of our knowledge, Yang et al.~\cite{yang2021multi} were the first to study the automatic smart contract comment generation problem.
Their proposed approach MMTrans learns the smart contract code representation from two heterogeneous modalities (i.e., SBT sequences and graphs based on abstract syntax trees). Their experimental results show that MMTrans can outperform some state-of-the-art baselines for source code summarization (such as Hybrid-DeepCom~\cite{hu2020deep}, code+gnn+GRU~\cite{leclair2019neural}, and Vanilla-Transformer~\cite{ahmad2020transformer}).
Since code reuse is common in smart contract development, we~\cite{yang2022ccgir} further propose a simple but effective information retrieval-based approach CCGIR. For the target smart contract code, this approach can effectively retrieve the most similar code from the historical repository and directly reuse the corresponding comment. In our empirical study, we find CCGIR can outperform different information retrieval approaches (such as NNGen~\cite{liu2018neural}) and MMTrans~\cite{yang2021multi}.

However, the performance of the previous studies~\cite{yang2021multi,yang2022ccgir} is still limited due to the amount of available training data. This issue is more obvious for the information retrieval-based approach CCGIR~\cite{yang2022ccgir}. 
To alleviate this problem, we aim to generate smart contract comments by using LLMs. To achieve this goal, we use the customized demonstration selection strategy to select high-quality demonstrations and use in-context learning to fully utilize the related knowledge in the LLMs for the target smart contract code.
To the best of our knowledge, we are the first to apply LLMs to smart contract comment generation. Our experimental results confirm that this direction is practical and feasible.

\subsection{Applying LLMs to Software Engineering Tasks}

Large language models (LLMs) refer to a class of artificial intelligence models that use an enormous amount of parameters and
are designed to process and generate human-like text based on large-scale language datasets~\cite{zhao2023survey}.
As a result, LLMs have been used for many mainstream software engineering tasks.
For the task of automated program repair, Xia et al.~\cite{xia2022less} utilized LLMs to directly generate correct code given the prefix
and suffix context. Then they~\cite{xia2023automated} conducted extensive empirical evaluations by considering nine different LLMs on five popular program repair datasets. Recently, they~\cite{xia2023keep} further leveraged test failure information and earlier patch
attempts in a conversational manner, which can prompt LLMs to generate more correct patches. 
For automated code generation tasks, Dong et al.~\cite{dong2023self} proposed a self-collaboration approach for code generation by ChatGPT. Liu et al.~\cite{liu2023improving} guided ChatGPT to generate better code with prompt engineering for two code generation tasks (i.e., text-to-code generation and code-to-code generation).
Liu et al.~\cite{liu2023your} proposed a code generation benchmarking framework, which can rigorously evaluate the functional correctness of the codes generated by ChatGPT.
For the task of source code summarization, Zhu et al.~\cite{zhu2023deep} performed empirical studies between deep learning methods (including LLMs) and information retrieval methods.
Sun et al.~\cite{sun2023automatic} performed source code summarization via ChatGPT and discussed the advantages and disadvantages of ChatGPT in this task.
In our study, we aim to apply LLMs to a new software engineering task and propose a novel approach {\tool}, which can effectively utilize the related domain knowledge in LLMs for smart contract comment generation via in-context learning.

\section{Conclusion}
\label{sec:conclusion}

In this study, we are the first to automatically generate smart contract comments by LLMs and in-context learning.
In our proposed approach, we utilize the customized demonstration selection strategy to select high-quality demonstrations, which can effectively utilize the related knowledge in LLMs via in-context learning for smart contract comment generation.
Our experimental results show {\tool} can significantly outperform baselines in automatic evaluation and human evaluation.
Our ablation studies also provided guidelines for effectively using {\tool}.

In the future, we first aim to further improve the performance of {\tool} by designing more effective demonstration selection strategies.
We second want to guide ChatGPT to generate higher-quality comments by prompt engineering.
Finally, we want to design more practical performance measures, which can effectively measure the semantic similarity between the ground-truth smart contract comments and the comments generated by {\tool}.


\section*{Acknowledgement}
The authors would like to thank the anonymous reviewers for their insightful comments and suggestions, which can substantially improve the quality of this work. 
Junjie Zhao and Xiang Chen have contributed equally to this work and they
are co-first authors.
This work is supported in part by the National Natural Science Foundation of
China (Grant no  62202419).

.

\section*{Declaration of Competing Interests}
The authors declare that they have no known competing financial interests or personal relationships that could have appeared to influence the work reported in this paper.
	
\section*{CRediT Authorship Contribution Statement}

\textbf{Junjie Zhao:} Data curation, Software, Validation, Conceptualization, Methodology, Writing -review \& editing.
\textbf{Xiang Chen:} Conceptualization, Methodology, Writing -review \& editing, Supervision.
\textbf{Guang Yang}: Conceptualization, Data curation, Software.
\textbf{Yiheng Shen:} Conceptualization, Validation.

\bibliography{mylib}
\bibliographystyle{elsarticle}

\vspace{1cm}

\noindent\textbf{Junjie Zhao} is currently pursuing the Master degree at the School of Information Science and Technology, Nantong University. His research interests include software repository mining.
\par
\vspace{1cm}
  
\noindent\textbf{Xiang Chen} received the B.Sc. degree in the school of management from Xi'an Jiaotong University, China in 2002. Then he received his M.Sc., and Ph.D. degrees in computer software and theory from Nanjing University, China in 2008 and 2011 respectively. 
He is currently an Associate Professor at the Department of Information Science and Technology, Nantong University, Nantong, China. He has authored or co-authored more than 120 papers in refereed journals or conferences, such as IEEE Transactions on Software Engineering, ACM Transactions on Software Engineering and Methodology, Empirical Software Engineering, Software Testing, Verification and Reliability, Information and Software Technology, Journal of Systems and Software, IEEE Transactions on Reliability, Journal of Software: Evolution and Process, Software - Practice and Experience, Automated Software Engineering, International Conference on Software Engineering (ICSE), The ACM Joint European Software Engineering Conference and Symposium on the Foundations of Software Engineering (ESEC/FSE), International Conference Automated Software Engineering (ASE), International Conference on Software Maintenance and Evolution (ICSME), International Conference on Program Comprehension (ICPC), and International Conference on Software Analysis, Evolution and Reengineering (SANER). His research interests include software engineering, in particular software testing and maintenance, software repository mining, and empirical software engineering. He received two ACM SIGSOFT distinguished paper awards in ICSE 2021 and ICPC 2023. He is the editorial board member of Information and Software Technology. More information about him can be found at: 

https://smartse.github.io/index.html.
\par
\vspace{1cm}

\noindent\textbf{Guang Yang} received the B.S. degree from Nantong University in 2019 and the M.Sc. degree from Nantong University in 2022. Now he is working on his Ph.D. degree at Nanjing University of Aeronautics and Astronautics. He has authored or co-authored more than 10 papers in refereed journals or conferences, such as ACM Transactions on Software Engineering and Methodology, Empirical Software Engineering, Journal of Systems and Software, Knowledge-based Systems, IEEE International Conference on Software Analysis, Evolution and Reengineering (SANER), International Conference on Software Maintenance and Evolution (ICSME), Asia-Pacific Symposium on Internetware (Internetware), International Conference on Software Engineering and Knowledge Engineering (SEKE), and Asia-Pacific Conference on Software Engineering (APSEC). His research interests include software engineering, especially automatic code generation and empirical software engineering.
\par
\vspace{1cm}

\noindent\textbf{Yiheng Shen} is currently pursuing the Master degree at the School of Information Science and Technology, Nantong University. His research interests include software repository mining.
\par

\end{document}